\documentclass[aps,pre,notitlepage,twocolumn,superscriptaddress,10pt]{revtex4-1}

\usepackage[dvipsnames]{xcolor}
\usepackage{amsmath}
\usepackage[pdftex]{graphicx}
\usepackage{subfigure}
\usepackage{lipsum}
\usepackage{lineno}

\begin{document}

\title{Herding and idiosyncratic choices: Nonlinearity and aging-induced transitions in the noisy voter model}

\author{Oriol Artime}
\affiliation{Instituto de F\'isica Interdisciplinar y Sistemas Complejos IFISC (CSIC-UIB), Campus UIB, 07122 Palma de Mallorca, Spain}
\affiliation{Fondazione Bruno Kessler, Via Sommarive 18, 38123 Povo (TN), Italy}
\author{Adri\'an Carro}
\affiliation{Institute for New Economic Thinking at the Oxford Martin School, University of Oxford, OX2 6ED Oxford, UK}
\affiliation{Mathematical Institute, University of Oxford, OX2 6GG Oxford, UK}
\author{Antonio F. Peralta}
\affiliation{Instituto de F\'isica Interdisciplinar y Sistemas Complejos IFISC (CSIC-UIB), Campus UIB, 07122 Palma de Mallorca, Spain}
\author{Jos\'e J. Ramasco}
\affiliation{Instituto de F\'isica Interdisciplinar y Sistemas Complejos IFISC (CSIC-UIB), Campus UIB, 07122 Palma de Mallorca, Spain}
\author{Maxi San Miguel}
\affiliation{Instituto de F\'isica Interdisciplinar y Sistemas Complejos IFISC (CSIC-UIB), Campus UIB, 07122 Palma de Mallorca, Spain}
\author{Ra\'ul Toral}
\affiliation{Instituto de F\'isica Interdisciplinar y Sistemas Complejos IFISC (CSIC-UIB), Campus UIB, 07122 Palma de Mallorca, Spain}

\begin{abstract}
We consider the herding to non-herding transition caused by idiosyncratic choices or imperfect imitation in the context of the Kirman Model for financial markets, or equivalently the Noisy Voter Model for opinion formation. In these original models this is a finite size transition that disappears for a large number of agents. We show how the introduction of two different mechanisms make this transition robust and well defined. A first mechanism is nonlinear interactions among agents taking into account the nonlinear effect of local majorities. The second one is aging, so that the longer an agent has been in a given state the more reluctant she becomes to change state.
\end{abstract}
\maketitle
\section{Introduction}
\label{sec:introduction}
A number of mechanistic models similar to those studied in Statistical Physics have been considered and analyzed to describe collective social phenomena \cite{ball2004critical,castellano2009statistical,fortunato2013JStatPhys,chakrabartiSociophysics,schweitzer2018sociophysics}. It turns out that, in some cases, models originally proposed in the Social Sciences are isomorphic to Statistical Physics models, as for example the famous segregation model of Schelling \cite{schelling1971dynamic} and a version of the Kinetic Ising model with exchange dynamics and vacancies \cite{Gauvin2009Schelling}. Such simple models of social behavior are sometimes dismissed as simplistic and the issue is what can be learnt from them. Following the physicists tradition of starting simple and considering science as the art of abstraction, these models aim to provide an understanding of mechanisms as a necessary step toward understanding data. Understanding means to explain why, which is about identifying mechanisms to establish cause-effect relations that go beyond inference from statistical correlations. It is also about unveiling underlying mechanisms beyond the description of collective patterns of behavior. The point of modeling is thus to understand data, rather than simply reproduce data. In this context, the purpose of a simple model is, essentially, i) to be able to pose a well defined question, ii) to isolate a mechanism of social interaction and determine emergent consequences at the collective level, iii) to establish cause-effect relations and iv) to check common sense wisdom \cite{watts2011obvious}.

A paradigmatic example of those simple models is the Voter Model, introduced first in biological \cite{clifford1973model} and mathematical statistics contexts \cite{holley1975ergodic,liggett2013stochastic} and considered in physics \cite{marro2005nonequilibrium} as a nonequilibrium lattice model. This model isolates imitation as a basic mechanism of social learning. The mathematical implementation of imitation, or herding, considers agents with a binary option (or state) that adopt the state of one of their neighbors chosen at random in their network of social interactions. The question is: If the only mechanism of interaction is pairwise imitation, when is agreement reached collectively, or when does coexistence of the two states persist? What we learn is that the answer to this question depends on the effective dimensionality of the network \cite{suchecki2005voter}. For regular lattices of dimension $d\leq2$ the system orders and dynamically approaches one of the two absorbing states of consensus with all agents adopting the same state. However, for larger effective dimension, which is the case of most complex networks, the system remains in a dynamical state of coexistence of states with a lifetime that diverges with system size. This result was considered counterintuitive \cite{castellano2003incomplete} since common sense might indicate that global scale interactions, obtained by introducing long range links in a regular square lattice of interactions (a small world network), should favor global consensus. However, the result is that long range links sustain a dynamical coexistence of states.

Social changes at the macroscale are aimed to be described by phase transitions occurring in these models for a critical value of a control parameter. In this way these changes can be traced back to the mechanism taken into account by the model. The Voter Model has no parameters, and therefore no possible phase transition. Kirman introduced \cite{kirman1993ants} a famous model used to describe herding behavior in financial markets with optimistic (buying) and pessimistic (selling) agents. The model accounts for fast switches between global optimistic and pessimistic states, but it also displays a transition from this global switching herding state to a non-herding state of dynamical coexistence with similar proportions of optimistic and pessimistic agents. In fact, Kirman's model is just a modification of the voter model, known as the noisy voter model \cite{carro2016noisy,peralta2018stochastic,Khalil:2018} and which has also been studied independently in many other contexts \cite{granovsky1995noisy,lebowitz1986percolation,fichthorn1989noise,considine1989comment}. The additional mechanism taken into account in Kirman's model is that of imperfect imitation: The change of state of an agent is not always determined by herding or imitation, but there are also idiosyncratic changes, i.e., decided by the agent independently of the choices of the other agents. This introduces a parameter in the model, the ratio of herding to idiosyncratic changes, so that a transition occurs for a critical value of this parameter. Imperfect imitation in the Voter Model has also been invoked to validate a metapopulation Voter Model against data from US presidential elections \cite{fernandez2014voter}.

The transition found in Kirman's model is a finite size transition: It depends on the number of agents, and it disappears in the limit of an infinite number of agents. In that limit there is always coexistence of the two states. From a Statistical Physics point of view this is not a proper phase transition since it does not exist in the thermodynamic limit. In fact, the model is not structurally stable and perturbations of the model lead to different results. Some structural modifications have been proposed so that the transition is no longer a finite size effect \cite{alfarano2005estimation,alfarano2008time}. In this paper we review how accounting for additional mechanisms makes the transition of Kirman's model robust and well defined. In particular, we consider separately the effect of nonlinear interactions and that of aging in the noisy voter model. Nonlinear interactions mean that the herding or imitation process is not implemented through dyadic agent-agent interactions, but that the change of state of a given agent is determined by group interaction with a set of her neighbors \cite{castellano2009nonlinear,peralta2018analytical}. By aging \cite{artime2018aging} we mean that the rate at which agents update their state is not constant, but rather agents have an internal clock measuring a persistence time, so that, the longer the agent remains in a given state, the less probable it is to update the state. We show that either of these additional mechanisms, when included in the noisy voter model, lead to a well defined transition between a herding and a non-herding state that occurs for a critical value of the ratio of herding to idiosyncratic changes, and that the transition continues to exist in the limit of a large number of agents.

The outline of the paper is as follows: In Section II we review the noisy voter model. Section III contains the discussion of nonlinear interactions and Section IV discusses the effect of aging. Some general conclusions are outlined in the final section.

\section{The noisy voter model}
\label{sec:the-noisy-voter-model}

We introduce next the standard noisy voter model, defining the notation and concepts used throughout the article, and we discuss as well the model's main characteristics. The properties derived here will be later used as a base case for comparison with those of the model variations, namely, its nonlinear version (Sec.~\ref{sec:nonlinear-noisy-voter-model}) and its version with aging (Sec.~\ref{sec:aging-in-the-noisy-voter-model}).

Let us consider a set of $ N $ elements, called ``agents''. They are placed on the nodes of a network of interactions characterized by the adjacency matrix $ A $, whose elements are $ A_{ij} = 1 $ if a connection between nodes $ i $ and $ j $ exists, and $ A_{ij} = 0 $ otherwise. The nature of these links is left here unspecified, but they are generally assumed to be interactions of some sort. The degree $k_i = \sum_j A_{ij}$ is defined as the total number of connections of node $i$. Beyond topological properties, each agent is endowed with a binary variable, its state or opinion, denoted by $ s_{i} \in \{0,1\} $. While the precise meaning of this binary variable does not concern us in this paper, typical interpretations include stock market traders holding an optimistic/pessimistic opinion about the future evolution of prices~\cite{kirman1993ants}, speakers using language A/B in bilingual societies~\cite{abrams2003linguistics,vazquez2010agent}, or, more generally, individuals holding one of two alternative opinions on a given topic. In order to characterize the global state of the system we introduce the global variable $ n(t) = \sum_i s_i$, defined as the total number of agents in state $ 1 $ at time $ t $. An alternative variable used for the macroscopic description is the intensive \textit{magnetization} $ m(t) = 2n/N - 1 $, defined in the interval $ [-1,1] $.

The dynamics of the noisy voter model defines how the transition in the states occur. The individual opinion of an agent $ s_i $ can change either by the effect of noise, in the form of random or idiosyncratic flips, i.e., independently of the opinions held by the agent's neighbors, or by pairwise copy or herding interactions with one of $ i $'s neighbors, the so-called voter update. The microscopic transition rates for each agent $ i $ with $ k_i $ neighbors read
\begin{equation}
 \begin{aligned}
 \label{eq:rates}
 \omega^+_i \equiv \omega \left( s_i=0 \to s_i=1 \right) &= \frac{a}{2} + \frac{(1 - a)}{k_i} \sum_{j = 1}^N A_{ij} s_j \, ,\\[5pt]
 \omega^-_i \equiv \omega \left( s_i=1 \to s_i=0 \right) &= \frac{a}{2} + \frac{(1 - a)}{k_i} \sum_{j = 1}^N A_{ij} (1 - s_j) \, .
 \end{aligned}
\end{equation}
In an all-to-all connected topology, the individual fractions of agents in one or the other state can be replaced by the global fractions. To obtain the global rates $\omega^{\pm}(n)$ one just needs to multiply by $ N - n $ and $ n $ the first and the second equations, respectively. , leading to
\begin{equation}
 \begin{aligned}
 \label{eq:all-to-all-rates}
 \omega^+(n)\equiv\omega(n\to n+1) & =(N-n)\left( \frac{a}{2} + (1-a) \frac{n}{N}\right) \, ,\\[5pt]
 \omega^-(n)\equiv\omega(n\to n-1) & = n\left(\frac{a}{2} + (1-a) \frac{N-n}{N}\right) \, .
 \end{aligned}
\end{equation}
The first term on the right-hand side, $a/2$, is the contribution of the noise. At each update event of agent $ i $, the noisy update is chosen with probability $ a  \in [0,1] $, resulting in a final state $ 1 $ half of the times and state $ 0 $ the other half, regardless of the former state. The voter update, the second term on the right-hand side, is performed with the complementary probability $ 1 - a $. In this case agent $ i $ chooses a random neighbor and blindly adopts the opinion of that neighbor. Thus, the fraction $ a/(1-a)$ stands for the relative ratio between noisy updates and voter updates and gives an idea of the strength or intensity of the noise.

\begin{figure*}
 \minipage{0.48\textwidth}
 \includegraphics[width=\linewidth]{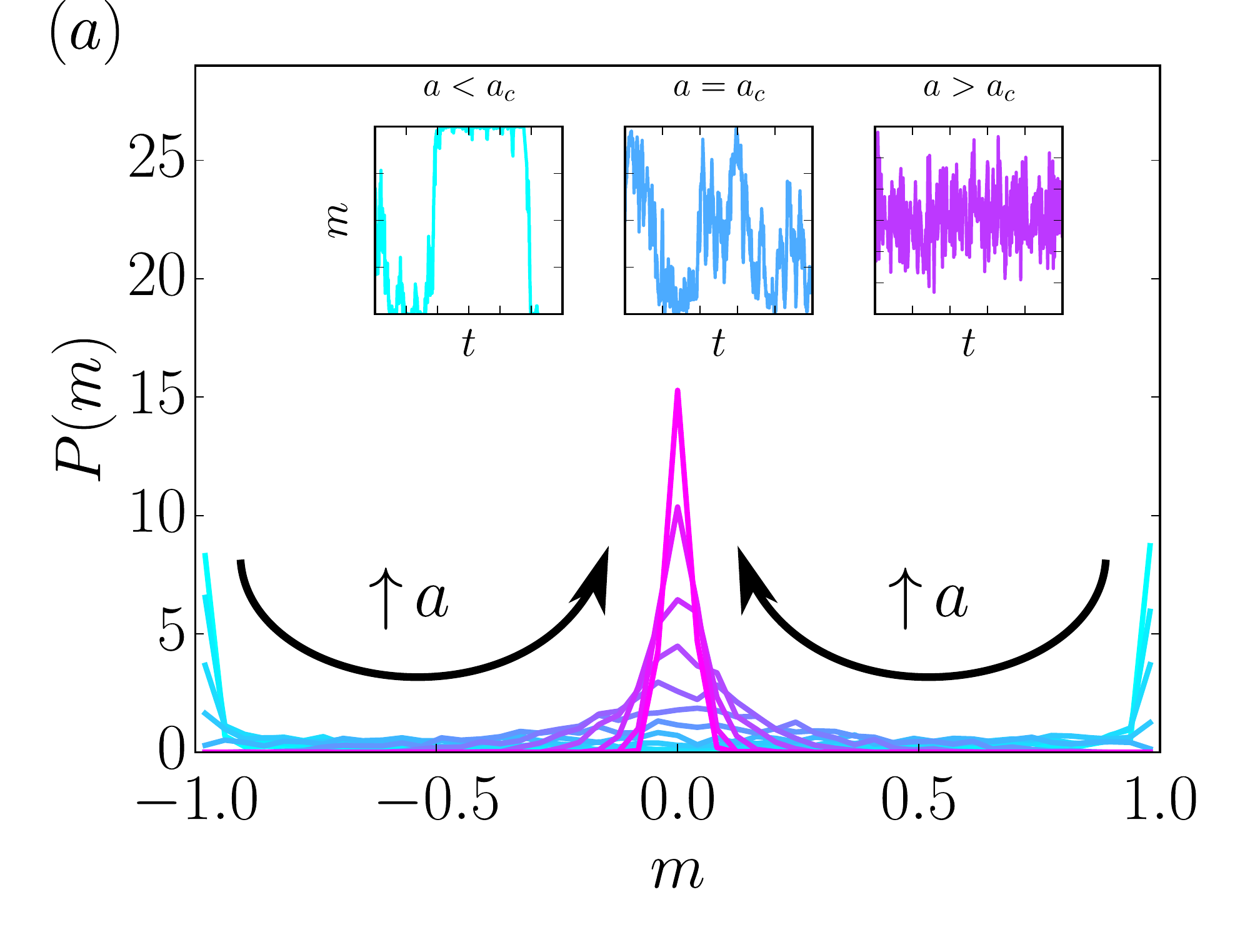}
 \endminipage\hfill
 \minipage{0.48\textwidth}
 \includegraphics[width=\linewidth]{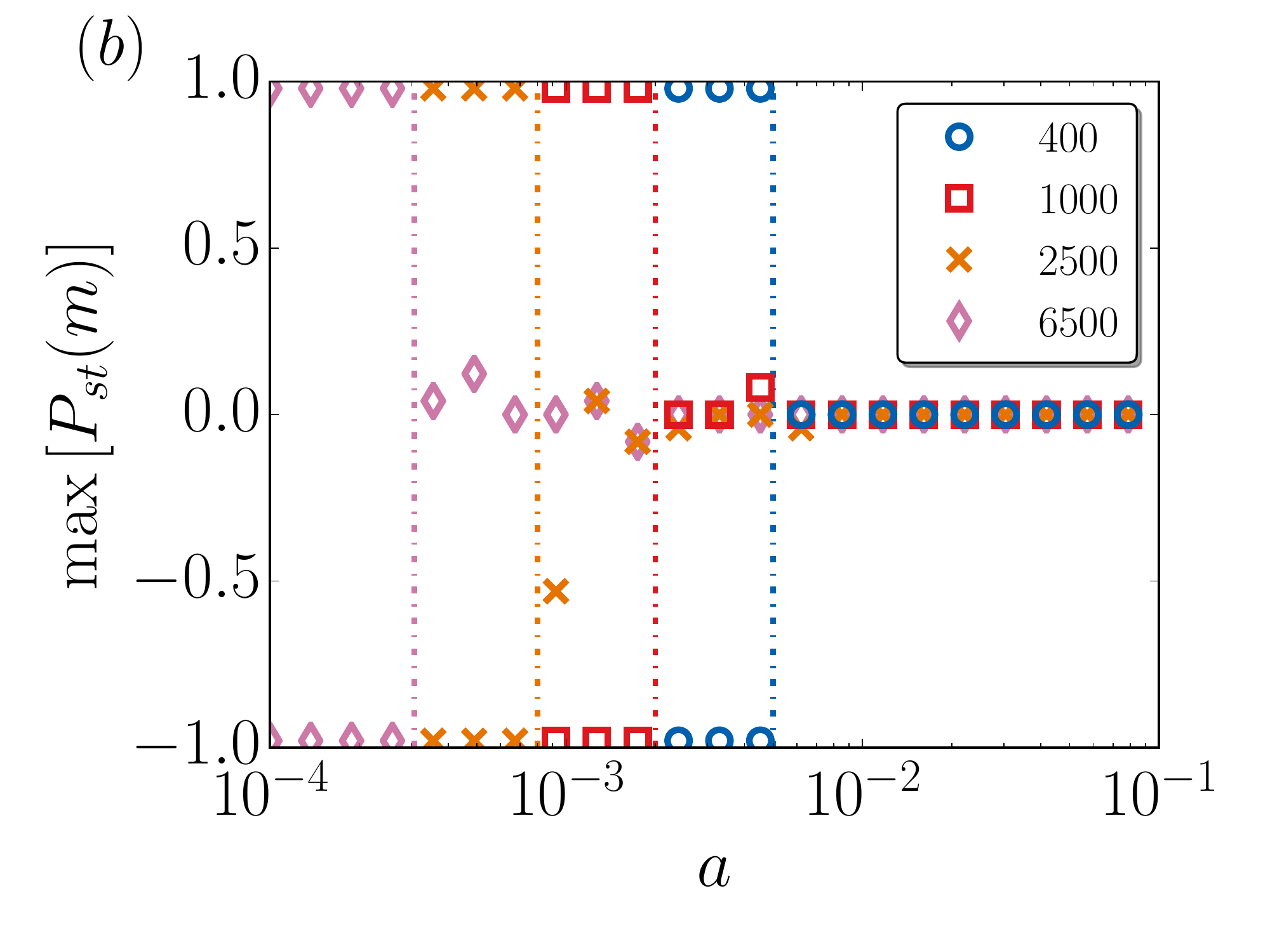}
 \endminipage
 \caption{Numerical simulations of the noisy voter model. In $(a)$, we show, for a system size $ N = 2000 $, the stationary probability density function (pdf) of the magnetization in the three different regimes. The insets show one typical trajectory of the dynamics in each of the regimes ($ a = 0.0005 $, $0.001$ and $0.05$). The considered range of values of $m$ in the insets is always $ [-1,1] $. In $(b)$, the maximum/maxima of the pdf as a function of the noise parameter $ a $ for different system sizes $N=400,1000,2500,6500$ as indicated. The vertical lines are the predictions for the critical noise for the different sizes considered in the plot.}\label{fig:Fig1}
\end{figure*}

Starting from the corresponding master equation of this stochastic process and using the rates Eqs.(\ref{eq:all-to-all-rates}), a standard approach \cite{VanKampen:2007,Toral-Colet:2014}, allows one to obtain a Fokker-Planck equation for the probability density function $ P(m,t)$ in the all-to-all connected topology,
\begin{equation}
 \label{eq:fokker-planck}
 \frac{\partial P(m,t)}{\partial t} = - \frac{\partial }{\partial m} \left[ F(m) P(m,t)\right] +
  \frac{\partial^2 }{\partial m^2} \left[ D(m) P(m,t) \right],
\end{equation}
with the drift term given by $ F(m) = - a m $ and the diffusion term by \mbox{$D(m) = [a + (1-a)(1-m^2)]/N$}. The negative sign and the odd dependence in the magnetization of the drift term indicate that the system has a tendency toward the state of equal coexistence of opinions $ m = 0 $, and this tendency is stronger the farther away from the coexistence. The diffusion function is formed by two terms with different effects on the dynamics of the system. The first one, independent of $ m $ and directly proportional to the noise $ a $, translates into state-independent stochastic fluctuations due to the ability of the agents to randomly change their state. The second term of the diffusion function, proportional to \mbox{$ (1 - a) $}, is a multiplicative noise term, i.e., a noise whose intensity depends on the state variable itself. In particular, this multiplicative noise is maximum for \mbox{$ m = 0 $} and vanishes for \mbox{$ m = \pm 1 $}, and so it statistically tends to move the system away from the center and toward the boundaries of the state variable $ m $, by making any random partial agreement on one or the other possible opinions stochastically grow toward a complete consensus, with all the agents sharing the same opinion. In the limiting case of vanishing noise $ a = 0 $ we recover the classical voter model \cite{liggett2013stochastic}, characterized by a zero drift and a vanishing diffusion term in the boundaries of the interval $ m= \pm 1 $. These are signatures of absorbing states: Once the system reaches there it cannot leave. As far as $ a \neq 0 $, consensus states exist but they are no longer absorbing.

The Fokker-Planck equation of the noisy voter model can be solved explicitly \cite{Artime:2018b, alfarano2006agent}, but we limit ourselves to an analysis of the stationary distribution $P_\text{st}(m)$, which can be found, after setting the time derivative to $ 0 $ in Eq.~\eqref{eq:fokker-planck}, in the form of an exponential function,
\begin{equation}
 \label{eq:fokker-planck-solution}
 P_\text{st}(m) = {\cal Z}^{-1} \cdot \exp \left[ -V(m) \right] \, ,
\end{equation}
where $\cal Z$ is a normalization constant and $V(m)$ is a ``potential function'' given by
\begin{equation}
 \label{eq:potential}
 V(m) = \int^{m} \frac{-F(z)+D'(z)}{D(z)} dz \, .
\end{equation}
Which, in our particular case, leads to
\begin{equation}
 \label{eq:stationary-distribution}
 P_\text{st} (m) = \mathcal{Z}^{-1} \left[ 1 + (a-1)m^2 \right]^\frac{2-a(N+2)}{2(a-1)},
\end{equation}
with the normalization constant being $\mathcal{Z} = 2 _2F_1 (\frac{1}{2}, 1 + \frac{aN}{2(a-1)}, \frac{3}{2}, 1-a) $, where $ _2F_1 $ is the hypergeometric function. The exponent of Eq.~\eqref{eq:stationary-distribution} determines the qualitative behavior of the system. When negative $ \left[a > 2/(N+2) \right] $, the stationary solution is a symmetric concave function with one peak at $ m = 0 $, the most disordered state. In this case the system is in the unimodal regime. If the exponent is positive $ \left[a < 2/(N+2) \right] $ the stationary state is found to be bimodal, i.e., a convex function with peaks at the consensus states $ m = \pm 1 $, the most ordered states. The lower the noise intensity, the higher the probability to find the system close to consensus. At the precise value where the exponent is $ 0 $ $ \left[a_c = 2/(N+2) \right] $ the stationary distribution is flat, meaning that the system does not have a preferred configuration, so any value of the magnetization is equiprobable. These three regimes are depicted in Fig.~\ref{fig:Fig1}$(a)$.

The nature of the transition from bimodal to unimodal distributions is discontinuous, in the sense that when the system crosses $ a_c $ the most probable state changes abruptly, from the borders to the middle of the magnetization space [see Fig.~\ref{fig:Fig1}$(b)$]. Another interesting feature of the transition of the noisy voter model is its size dependence. In a real system we encounter finite populations, thus the transition will be always observable. However, to characterize a bona-fide phase transition one needs to take the thermodynamic limit $ N \to \infty $, where the thermodynamic quantities suffer actual discontinuities and/or divergences. In the case of the noisy voter model, the transition point vanishes $a_c$ when the infinite size limit is taken. In other words, there is no bimodal regime in this limit, i.e., no possible change of phase. This scenario resembles the one-dimensional Ising model, with its trivial phase transition at zero temperature \cite{yeomans1992statistical}.

With the purpose of keeping the bimodal regime accessible even in the thermodynamic limit, and thus of avoiding the finite-size character of the transition in the original noisy voter model, an alternative formulation of the voter rule within the transition rates at Eq.~\eqref{eq:rates} has been proposed in the literature \cite{alfarano2005estimation,alfarano2008time,carro2015markets}. In particular, an interaction mechanism that is, for each agent, proportional to the absolute number of agents in the opposite state ---rather than the corresponding fraction--- has been proposed. This is tantamount to replacing $1-a$ by $N(1-a)$ in the rates (\ref{eq:all-to-all-rates}). In this case, the critical point of the transition becomes size-independent, $a_c=2/3$, and thus the transition remains in the thermodynamic limit. It should be note, however, that this redefinition of the voter rule rests on the assumption that agents change opinion depending only on the number of agents with the opposite opinion, completely disregarding the number of neighbors with their same opinion. For this reason, we consider the original formulation of the voter update rule to be of more general application, and thus we will use the transition rates as expressed in Eq.~\eqref{eq:rates}.

Several recent contributions have confirmed the existence of the referred finite-size transition when the all-to-all assumption is relaxed and the system is embedded in a more complex network topology. In particular, three different analytical approaches have been proposed to explain the effect of the network of interactions on the results of the noisy voter model. A mean-field approximation was initially proposed in the literature \cite{alfarano2009network,diakonova2015noise}, leading to an analytical solution for the critical point which coincides with the all-to-all case, \mbox{$ a_c = 2/(N+2) $}, and is thus independent of any property of the network other than its size. A second approach has been recently introduced based on an annealed approximation for uncorrelated networks \cite{carro2016noisy}, allowing to deal with the network structure as parametric heterogeneity. This study has shown that the variance of the underlying degree distribution has a strong influence on the location of the critical point of the transition, namely leading to larger critical points the larger the degree heterogeneity of the underlying network. Finally, a third approach has been presented in a recent contribution \cite{peralta2018stochastic} based on a full stochastic description of the pair approximation scheme to study binary-state dynamics on heterogeneous networks. This study has confirmed the role played by the variance of the degree distribution, as well as improved on the accuracy of previous methods. In all these theoretical approaches, and coincident with the results of numerical simulations, the critical point $a_c$ tends to zero as the system size goes to infinity.

Complex networks are characterized by a heterogeneous degree distribution such that agents are connected to a different number of neighbors. Another line of research has considered the role that the {\sl intrinsic} heterogeneity of the agents can have on the transition point. Intrinsic heterogeneity refers to the fact that different agents can have different levels of noise or herding, reflecting the different relevance that these two terms can have in an individual's behavior. A general framework for binary state models, that is also useful to handle network heterogeneity, has been developed in \cite{Lafuerza:2013}. The result is that, although intrinsic heterogeneity can alter the location of the transition point, its value still scales with the inverse size and, therefore, the transition disappears in the thermodynamic limit. An extreme case of agents' heterogeneity is the presence of zealots, or agents that never change their state \cite{Mobilia:2003}. The role of zealots in the noisy voter model within a mean-field approach has been considered in \cite{Khalil:2018} with the conclusion that the dynamics of a system of identical voters with the inclusion of some zealots, is equivalent to that of an intrinsic heterogeneous population without zealots and displays a rich phenomenology with the possible presence of asymmetric phases. However, the finite-size character of the transition remains in the presence of zealots.

In the following sections we will discuss different mechanisms that transform the finite-size transition of the noisy voter model into a robust, well-defined phase transition.

\section{Nonlinear noisy voter model}
\label{sec:nonlinear-noisy-voter-model}


In this section, we present a modification of the noisy voter model allowing for non-linear interactions between agents. As shown below, exact analytical solutions can be found for the main model variables in an all-to-all scenario. In particular, we present here solutions for the location of the maxima of the probability distribution of the magnetization as a function of the parameters of the model and the population size. As a consequence, we will be able to show and discuss the main features of the phase diagram characterizing the model. Finally, we will comment on how these results are modified by releasing the all-to-all constraint and allowing for more complex topologies.


As mentioned above, the canonical noisy voter model includes both a noise mechanism, in the form of random changes of state, and a herding mechanism, in the form of pairwise copy interactions between nearest neighbors in the network ---the so-called voter update rule. In particular, the dyadic character of the voter update rule, in which a node copies the state of a randomly selected neighbor, implies that the probability of a node changing its state due to this mechanism is proportional to the fraction of neighboring nodes in the opposite state. In general, however, interactions do not need to be dyadic and, indeed, in a number of social contexts, more complex mechanisms of group interaction seem to be in place. A more general way of modeling collective interaction is to assume that the associated probability to change state is proportional to a power $\alpha$ of the fraction of neighbors in the opposite state, thus allowing for nonlinear functional forms.


In this way, we rewrite the transition rates at Eq.~\eqref{eq:rates} as
\begin{equation}
 \begin{aligned}
 \label{eq:non-linear-rates}
 \omega^+_i &= \frac{a}{2} + (1 - a) \left( \frac{1}{k_i} \sum_{j = 1}^N A_{ij} s_j \right)^{\alpha} \, ,\\[5pt]
 \omega^-_i &= \frac{a}{2} + (1 - a) \left( \frac{1}{k_i} \sum_{j = 1}^N A_{ij} (1 - s_j) \right)^{\alpha} \, ,
 \end{aligned}
\end{equation}
such that the linear dependence corresponding to the random imitation in the traditional noisy voter model is easily recovered for $\alpha=1$. Values of $\alpha>1$ correspond to a probability of imitation below the random case, thus modeling individuals who are more resistant to follow the opinion of their neighbors holding the opposite state. As a consequence of this aversion to change, a larger fraction of neighbors in the opposite state is needed for the agent to switch her state, as compared to the purely random imitation of the traditional voter update rule. On the contrary, values of $\alpha<1$ correspond to a probability of imitation above the random case, thus making it easier for agents to follow the opinion of their neighbors holding the opposite state. As a consequence of this preference for change, even a small fraction of neighbors in the opposite state is likely to trigger the agent to switch her state, as compared to the purely random imitation case. While values of $\alpha>1$ have been found to fit the data in some problems of language competition~\cite{abrams2003linguistics}, values of $\alpha<1$ have been generally considered in social impact theory~\cite{nowak1990private}. Note that the cases where $\alpha$ is an integer, $\alpha=2,3,\dots$, are equivalent to a process in which an individual changes state if and only if, after checking the state of $\alpha$ randomly selected neighbors (allowing for repetitions in the selection), all of them happen to be in the opposite state to the one held by the individual~\cite{castellano2009nonlinear}. For the sake of generality and in order to be able to study the robustness of the transition of the noisy voter model to small perturbations around $\alpha=1$, we focus here on continuous values of $\alpha$. A version of this nonlinear model has been used recently in a model of active particles \cite{Escaff:2018}. In this version of the model the two states represent possible directions of the velocity of active particles in a one-dimensional line. There is a rate to switch the direction of movement that takes into account the noise and non-linear herding mechanisms that have been described before. The novelty in this model compared to the ones analyzed here is that the set of neighbors of a given active particle is not derived statically from a network structure, but the range of interaction is solely dictated by the physical distance between the particles and, hence, the number of possible interacting particles changes with time. Using an exponent $\alpha=2$, it is found that there is a transition to a flocking state in which particles move together in the same direction.


As mentioned above, in the all-to-all scenario, i.e., when each node is connected to all the other nodes, individual fractions of agents in one or the other state coincide with the global fractions. Furthermore, being all nodes equivalent, the only variable of interest is $n$ and global transition rates for the system as a whole can be easily obtained as
\begin{equation}
 \begin{aligned}
 \label{eq:non-linear-global-rates}
 \omega^+(n) &= (N - n) \left( \frac{a}{2} + (1 - a) \left( \frac{n}{N} \right)^{\alpha} \right) \, ,\\[5pt]
 \omega^-(n) &= n \left( \frac{a}{2} + (1 - a) \left( \frac{N - n}{N} \right)^{\alpha} \right) \, .
 \end{aligned}
\end{equation}
Following the procedure described in the previous section with these modified rates, a Fokker-Planck equation~\eqref{eq:fokker-planck} can be obtained for this nonlinear version of the noisy voter model. In particular, the drift and diffusion terms are respectively found to be
\begin{widetext}
\begin{equation}
 \begin{aligned}
 \label{eq:drift-and-diffusion}
 F(m) &= - a m + 2^{-\alpha} (1 - a) (1 - m^2)   \left( (1 + m)^{\alpha - 1} - (1 - m)^{\alpha - 1} \right)\, ,\\[5pt]
 D(m) &= \left[a + 2^{-\alpha} (1 - a) (1 - m^2)  \left( (1 + m)^{\alpha - 1} + (1 - m)^{\alpha - 1} \right)\right]/N \, .
 \end{aligned}
\end{equation}
\end{widetext}


The stationary distribution, obtained from by Eqs.(\ref{eq:fokker-planck-solution}-\ref{eq:potential}) using these drift and diffusion functions, is a complicated expression. However, for the sake of analyzing the phase diagram of the model, we only need to determine the number and location of the local maxima of the stationary distribution $P_\text{st}(m)$ as a function of the model parameters. Moreover, due to the exponential form of Eq.~\eqref{eq:fokker-planck-solution}, the conditions for a point \mbox{$m = m_{*}$} to be a local extreme of the stationary distribution $P_\text{st}(m)$ and for this extreme to be a maximum can be more simply written in terms of the potential function $V(m)$ as, respectively, \mbox{$V'(m_{*}) = 0$} and \mbox{$V''(m_{*}) < 0$}. Note, nevertheless, that these maxima can also be located at the boundary values \mbox{$m_{*} = \pm 1$}.


When imposing the first condition, \mbox{$V'(m_{*}) = 0$}, it is easy to notice that there is a trivial local extreme at \mbox{$m_{*} = 0$}, which corresponds to the perfectly balanced case ---with the same number of agents in each of the two possible states. Applying the second condition to understand if this point is a minimum or a maximum, it is useful to write the second derivative of the potential function $V(m)$ at \mbox{$m_{*} = 0$} as
\begin{equation}
 \label{eq:stability_zero}
 V''(m_{*} = 0) = \frac{2^{\alpha}}{2^{\alpha} \varepsilon + 1} ( \varepsilon - \varepsilon_c(N) ) \, ,
\end{equation}
where we have defined a noise-herding relative intensity parameter $\varepsilon \equiv a/(2(1 - a))$ and its critical value as a function of the system size $\varepsilon_c(N)$, i.e., the value of $\varepsilon$ at which the second derivative of the potential function at $m_{*} = 0$ changes sign, namely
\begin{equation}
 \label{eq:critical-epsilon}
 \varepsilon_c (N) = 2^{-\alpha} \left( \alpha - 1 + \frac{\alpha (3 - \alpha)}{N} \right) \, .
\end{equation}
According to these definitions, for \mbox{$\varepsilon < \varepsilon_c(N)$} then \mbox{$m_{*} = 0$} corresponds to a minimum of the stationary state probability distribution, while for \mbox{$\varepsilon > \varepsilon_c(N)$} it corresponds to a maximum.


In the thermodynamic limit of large system sizes, the critical value of this new relative intensity parameter behaves as
\begin{equation}
 \label{eq:critical-epsilon-large-N-limit}
 \varepsilon_c \equiv \varepsilon_c (\infty) = 2^{-\alpha} \left( \alpha - 1 \right) \, .
\end{equation}
As a consequence, in this limit, the change in sign in the second derivative of the potential function, and thus the transition between \mbox{$m_{*} = 0$} corresponding to a maximum and a minimum, occurs at a finite positive value of $\varepsilon$ for \mbox{$\alpha > 1$}. For \mbox{$\alpha < 1$}, on the contrary, since \mbox{$\varepsilon_c(\infty) < 0$} and $\varepsilon$ must be non-negative, then \mbox{$m_{*} = 0$} is always a maximum, and thus there is no transition. In the particular case of the canonical (linear) noisy voter model, i.e., for \mbox{$\alpha = 1$}, we have \mbox{$\varepsilon_c(N) = 1/N$}, and thus the transition is only a finite-size effect, such that, in the thermodynamic limit, the point \mbox{$m_{*} = 0$} corresponds to a maximum for all values of \mbox{$\varepsilon > 0$}.


Since for \mbox{$\alpha < 1$} there is no transition and the \mbox{$\alpha = 1$} case is known to be characterized by a unique finite-size transition between a unimodal and a bimodal probability distribution, we focus here on the \mbox{$\alpha > 1$} case when searching for other maxima apart from the trivial one at \mbox{$m_{*} = 0$}. Looking at Eq.~\eqref{eq:critical-epsilon-large-N-limit} and bearing in mind the definition of $\varepsilon$, we see that for \mbox{$\alpha > 1$} and in the thermodynamic limit both noise and herding, respectively $a/2$ and \mbox{$(1 - a)$}, are of a similar order of magnitude around the critical region close to the transition. As a consequence of this, and keeping also in mind the scaling with $N$ of the drift and diffusion terms  at Eq.~\eqref{eq:drift-and-diffusion}, we see that, in the thermodynamic limit, the term $D'(m)$ can be safely ignored with respect to $F(m)$ in the numerator of Eq.~\eqref{eq:potential}. It thus follows that the conditions for a point $m_{*}$ to be a local extreme of the stationary probability distribution and for this extreme to be a maximum can be rewritten in terms of the drift function $F(m)$ as, respectively, \mbox{$F(m_{*}) = 0$} and \mbox{$F'(m_{*}) > 0$}. Note that we have used, for the second condition, the fact that the first condition must be fulfilled and that the diffusion function is always positive.


Given that it is not possible to find a closed solution for the equation \mbox{$F(m_{*}) = 0$}, we propose here an expansion of the drift function in power-series of $m$ around \mbox{$m = 0$} and up to $O(m^7)$,
\begin{align}
 \label{eq:drift-expansion}
 F(m) &= 2 (1 - a) \left( \varepsilon_{c} - \varepsilon \right) m \\[5pt]
 &+ \frac{(1 - a)}{3} 2^{-\alpha} (\alpha - 5) (\alpha - 1) \alpha m^3 \notag\\[5pt]
 &+ \frac{(1 - a)}{15} 2^{-2-\alpha} (\alpha - 9) (\alpha - 3) (\alpha - 2) (\alpha - 1) \alpha m^5 \notag\\[5pt]
 &+ O(m^7) \, . \notag
\end{align}
Note that, for \mbox{$a = 0$}, this expression coincides with the derivative of the potential function previously found in the literature for the noiseless nonlinear voter model~\cite{vazquez2010agent} \footnote{Note that, due to a misprint, a factor \mbox{$(\alpha - 9)$} is missing at Eq.~(10) of Ref.~\cite{vazquez2010agent}.}. While the expansion at Eq.~\eqref{eq:drift-expansion} is, strictly speaking, only valid for \mbox{$m \approx 0$}, this approximation is enough to identify the existence and stability ---maximum or minimum character--- of any additional extrema, as well as the corresponding transitions.

\begin{figure}[ht]
 \centering
 \includegraphics[width=0.5\textwidth]{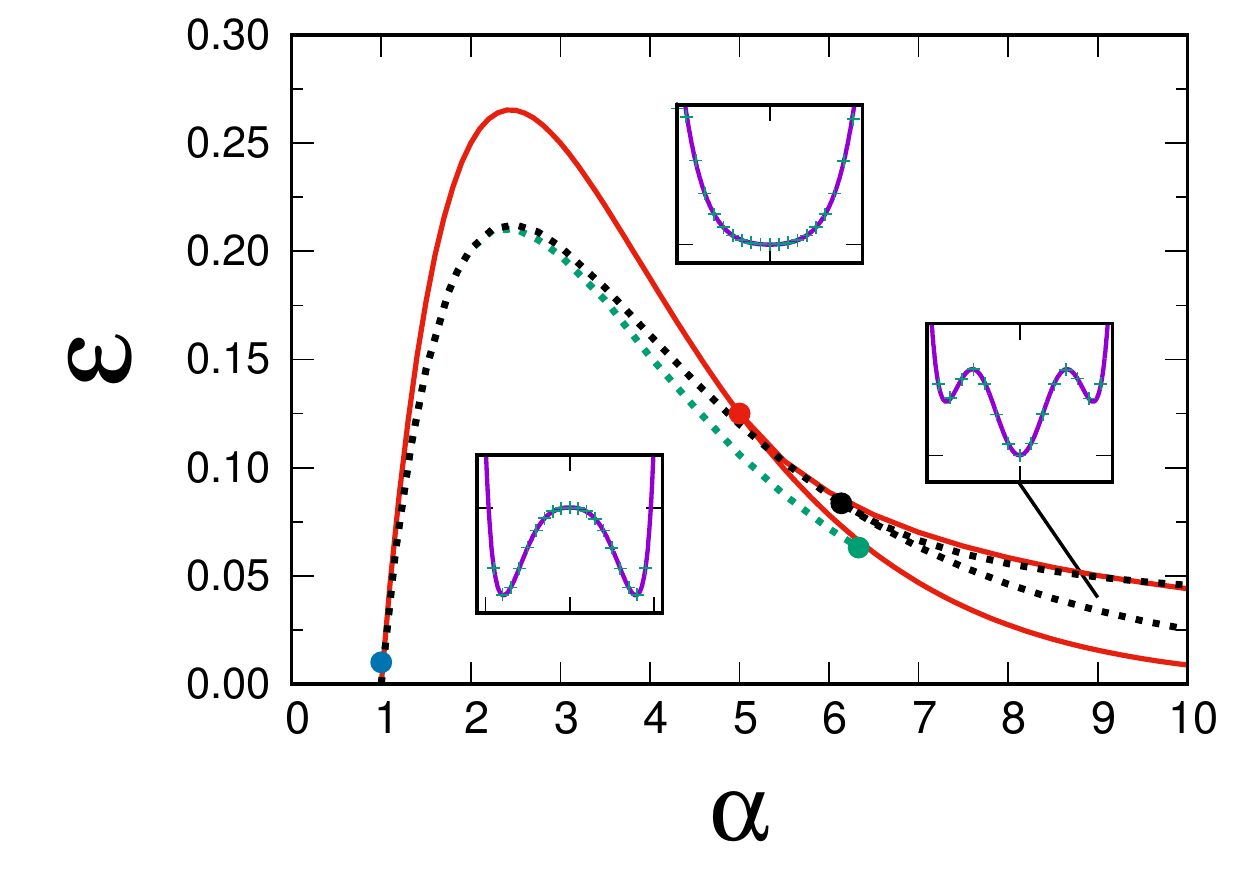}
 \caption{Phase diagram of the non-linear noisy voter model, with the shape of the potential $V(m)$ Eq.(\ref{eq:potential}) for the different phases of the model in the snapshots. The dot at \mbox{$\alpha = 1$} (in blue) is the finite-size transition point $\varepsilon_c(N)$, given by Eq.~\eqref{eq:critical-epsilon}, for a system size \mbox{$N = 100$}. The solid lines correspond to the all-to-all scenario, while the dotted lines correspond to a $15$-regular network (black) and the scale free network $P_{5 \leq k \leq 966} \sim k^{-2.34}$ with $\mu \approx 15$ (green). The trimodal region is delimited by the transition lines $\varepsilon_{c}$ and $\varepsilon_{t}$, which for the all-to-all scenario are given, respectively, by Eqs.~\eqref{eq:critical-epsilon} and~\eqref{eq:critical-epsilon-large-N-limit-2}, with a tricritical point (red) at \mbox{$\alpha = 5$, $\varepsilon = 1/8$}. The tricritical point for the $15$-regular network (black) is at \mbox{$\alpha = 6.14$, $\varepsilon = 0.084$} and for the scale free network (green) at \mbox{$\alpha = 6.33$, $\varepsilon = 0.063$}. For the sake of clarity, the figure does not show the trimodal region for the scale free network.}
 \label{fig:modes1}
\end{figure}


In particular, using the expansion at Eq.~\eqref{eq:drift-expansion} we find five solutions for the equation defining the extrema of the probability distribution, \mbox{$F(m_{*}) = 0$}. Namely, the trivial solution \mbox{$m_{*} = 0$} and other four (possibly complex) roots $\pm m_{*}^{+}$ and $\pm m_{*}^{-}$, derived from
 \begin{align}
 \label{extrema}
 (m_{*}^\pm)^2 \approx & \frac{10 (\alpha -5) }{(\alpha - 2) (\alpha - 3) (\alpha - 9)} \\[5pt]
 \times & \left[ \pm \sqrt{1 - \frac{6}{5} \frac{(\alpha - 2) (\alpha - 3) (9 - \alpha)}{\alpha (\alpha - 5)^2} \left( \frac{\varepsilon - \varepsilon_{c}}{\varepsilon_{c}} \right)} - 1 \right] \, . \notag
 \end{align}
Note that these four roots can be real or imaginary depending on the values of the parameters $\varepsilon$ and $\alpha$. In order to find the maxima of the probability distribution $P_\text{st}(m)$, however, we are only interested in values of $m_{*}$ which are real and inside the interval $[-1, 1]$. Thus, depending on the parameters, we have:
\begin{enumerate}
 \item In the region $1 < \alpha < 5$,
 \begin{itemize}
 \item for $\varepsilon < \varepsilon_{c}$ only the pair of solutions $\pm m_{*}^+$ are real and correspond to probability maxima,
 \item while for $\varepsilon > \varepsilon_{c}$ all four roots are imaginary.
 \end{itemize}
 Therefore, in this region, the line $\varepsilon_{c}$ separates a unimodal (single maximum of the probability distribution) from a bimodal (two real maxima) regime, see Fig.~\ref{fig:modes1}.
 \item In the region $\alpha > 5$, we can define a new critical value of the relative intensity parameter $\varepsilon$ as
 \begin{equation}
 \label{eq:critical-epsilon-large-N-limit-2}
 \varepsilon_{t} \approx \varepsilon_{c} \cdot \left[ 1 + \frac{5 \alpha (\alpha - 5)^2}{6 (\alpha - 2)(\alpha - 3)(9 - \alpha)} \right] \, ,
 \end{equation}
 such that
 \begin{itemize}
 \item for $\varepsilon < \varepsilon_{c}$ only the pair of solutions $\pm m_{*}^+$ are real and correspond to probability maxima,
 \item for $\varepsilon_{c} < \varepsilon < \varepsilon_{t}$ all four roots are real, $\pm m_{*}^-$ corresponding to probability maxima, and $\pm m_{*}^+$ to probability minima,
 \item and for $\varepsilon > \varepsilon_{t}$ all four roots are imaginary.
 \end{itemize}
 Therefore, in this region, the line $\varepsilon_{c}$ separates a bimodal from a trimodal regime, while the line $\varepsilon_{t}$ separates a trimodal from a unimodal regime, see Fig.~\ref{fig:modes1}. These two lines meet at the tricritical point given by \mbox{$\alpha = 5$} and \mbox{$\varepsilon = 2^{-3} = 0.125$}.
\end{enumerate}
Note that both the non-trivial roots at Eq.~\eqref{extrema} and the critical value $\varepsilon_{t}$ at Eq.~\eqref{eq:critical-epsilon-large-N-limit-2} are approximate expressions, as they are based on the expansion at Eq.~\eqref{eq:drift-expansion}. By means of numerical analysis ---using the exact forms for the potential function at Eq.~\eqref{eq:fokker-planck-solution} and the drift and diffusion functions at Eq.~\eqref{eq:drift-and-diffusion}---, we have verified that, first, these expressions are accurate only for \mbox{$\alpha < 7$}, and second, the presented classification of maxima and minima is completely general and additional extrema or transitions are not observed.


Let us now briefly comment on the specific character or continuity of the identified transitions:
\begin{enumerate}
 \item For $\alpha = 1$, we recover the canonical noisy voter model behavior, characterized by a discontinuous transition at $\varepsilon_c$. In particular, for finite size systems, the discontinuous transition separates a regime with two maxima at \mbox{$m_{*} = \pm 1$} for \mbox{$\varepsilon < \varepsilon_c(N)$} from a regime with a single maximum at \mbox{$m_{*} = 0$} for \mbox{$\varepsilon > \varepsilon_c(N)$}. In the thermodynamic limit, it is only at $\varepsilon = \varepsilon_c = 0$, i.e., for the noiseless voter model, that the probability distribution is a sum of two delta functions at \mbox{$m = \pm 1$}, leading to the maxima \mbox{$m_{*} = \pm 1$} if \mbox{$\varepsilon = 0$} and \mbox{$m_{*} = 0$} for all other \mbox{$\varepsilon > 0$}. Formally, the scaling of the transition can be written as \mbox{$|m_{*}| \sim (\varepsilon_{c}(N) - \varepsilon)^{\beta}$} with \mbox{$\beta = 0$}.
 \item For $1 < \alpha < 5$, the line \mbox{$\varepsilon = \varepsilon_{c}$} corresponds to a classical (Landau) second order (continuous) phase transition from a (relatively) ``ordered'' phase with \mbox{$m_{*} \ne 0$} to a ``disordered'' phase with \mbox{$m_{*} = 0$}. The scaling of the transition is \mbox{$|m_{*}| \sim (\varepsilon_c - \varepsilon )^{\beta}$} for \mbox{$\varepsilon \le \varepsilon_{c}$} with \mbox{$\beta = 1/2$} and \mbox{$|m_{*}| = 0$} for \mbox{$\varepsilon \ge \varepsilon_{c}$}.
 \item For \mbox{$\alpha = 5$}, the transition is still continuous, but with a scaling as \mbox{$|m_{*}| \sim (\varepsilon_c - \varepsilon)^{\beta}$} with \mbox{$\beta = 1/4$}.
 \item For \mbox{$\alpha > 5$}, an increase of $\varepsilon$ leads first to a first-order transition from a bimodal to a trimodal distribution at \mbox{$\varepsilon = \varepsilon_c$}, implying a discontinuity in the location of the absolute maximum from \mbox{$|m_{*}| > 0$} to \mbox{$m_{*} = 0$}. The local maxima at \mbox{$|m_{*}| > 0$} remain up to \mbox{$\varepsilon = \varepsilon_t$}, when they disappear and the distribution suddenly becomes unimodal, thus corresponding to a discontinuous transition.
\end{enumerate}
In this way, we see that the well-studied bimodal-unimodal finite-size transition of the noisy voter model in an all-to-all network turns into a classical second-order transition for \mbox{$1 < \alpha < 5$}, no transition for \mbox{$\alpha < 1$}, a finite-size transition for the traditional linear case \mbox{$\alpha = 1$}, and a couple of discontinuous first-order transitions, respectively between a bimodal and a trimodal regime and between a trimodal and a unimodal regime, for \mbox{$\alpha > 5$}.


By using of a pair approximation scheme, we have also studied the role of the network of interactions in the presented results of the non-linear noisy voter model \cite{peralta2018analytical}. The main effect of the complex network is to shift the transition lines and to modify the system-size dependence of the critical lines and points dependent on it. In particular, both (approximate) analytical and numerical results lead us to conclude that the critical point $\varepsilon_{c}$ is lowered with respect to the all-to-all solution, depending on the mean degree $\mu$ and some negative moments ($\gamma<0$) of the degree distribution, \mbox{$\mu_{\gamma}\equiv \langle k^\gamma\rangle$}. In the case of highly heterogeneous networks with degree distributions whose second moment $\mu_2$ experiences large changes with system size, such as scale-free networks \mbox{$\mu_2 \sim N^{b}$} with \mbox{$0 < b < 1$}, the $N$-dependent scaling laws are also modified by the network structure. Interestingly, this modification can be captured by the introduction of an effective system size that decreases with the degree heterogeneity of the network. In this way, the previously presented scaling laws for the all-to-all scenario are still valid for more complex networks by simply replacing the actual system size by the effective one. In particular, this effective system size can be defined as \mbox{$N_{\text{eff}} = \frac{\mu^2}{\mu_2} N \sim N^{1-b}$}. In the limit of highly heterogeneous networks results may be $N$-independent or with a very weak dependence on system size.


\section{Aging in the noisy voter model}
\label{sec:aging-in-the-noisy-voter-model}

In the previous section, we have shown the rich behavior that nonlinear interactions induce in the stationary magnetization distribution of the noisy voter model. These new interactions naturally modify the transition rates~\eqref{eq:non-linear-rates}. However, from the temporal point of view, the nonlinear noisy voter model remains completely Markovian. Still, non-Markovianity is known to arise in many real systems, specially in those where humans interact with each other. Incorporating this realistic phenomenon in the modeling framework is not always an easy task: The mathematical techniques to analyze these systems are not so well-established as those for Markovian dynamics, and the numerical simulations are usually more difficult to implement. In this section, we discuss the noisy voter model with aging. Aging is taken as the dependence of the transition rates between states on the age of the agents. In this context, ``age'' is understood as the time elapsed since the last change of state. We will show that, by including a particular age dependence in the rates of the noisy voter model, the finite-size nature of the transition is overcome and a continuous phase transition well-defined in the thermodynamic limit appears. Along this section, we characterize the transition by (i) finding an analytical expression for the stationary mean magnetization, (ii) discussing its universality class, and (iii) looking for new broken symmetries due to the temporal dimension.

The main modification is to assign an internal age variable $\tau=0,1,2,\dots$ to each agent of the population.  The age accounts for the time elapsed since the agent's last change of state. When a node modifies its state either due to noise or imitation, the age is reset to $ 0 $. Otherwise, $ \tau $ increases in one unit. Initially, all nodes start with $ \tau = 0 $. The noisy update is unaffected by aging (like the nonlinearity in the rates~\eqref{eq:non-linear-rates}). Regardless of the age, an agent is selected to update with probability $ a $ via the noisy process defined in Section II. The pairwise update, nevertheless, is altered. Before performing it, the selected node must be first activated with a probability $ p(\tau) = 1/ (\tau+2) $. Hence, the older a node, the more difficult to activate to change state with the voter mechanism. This specific form of aging is able to reproduce power-law inter-event time distributions \cite{fernandez2011update}, a feature encountered in many human-related systems \cite{karsai2018bursty}.

\begin{figure}
 \includegraphics[width=0.8\linewidth]{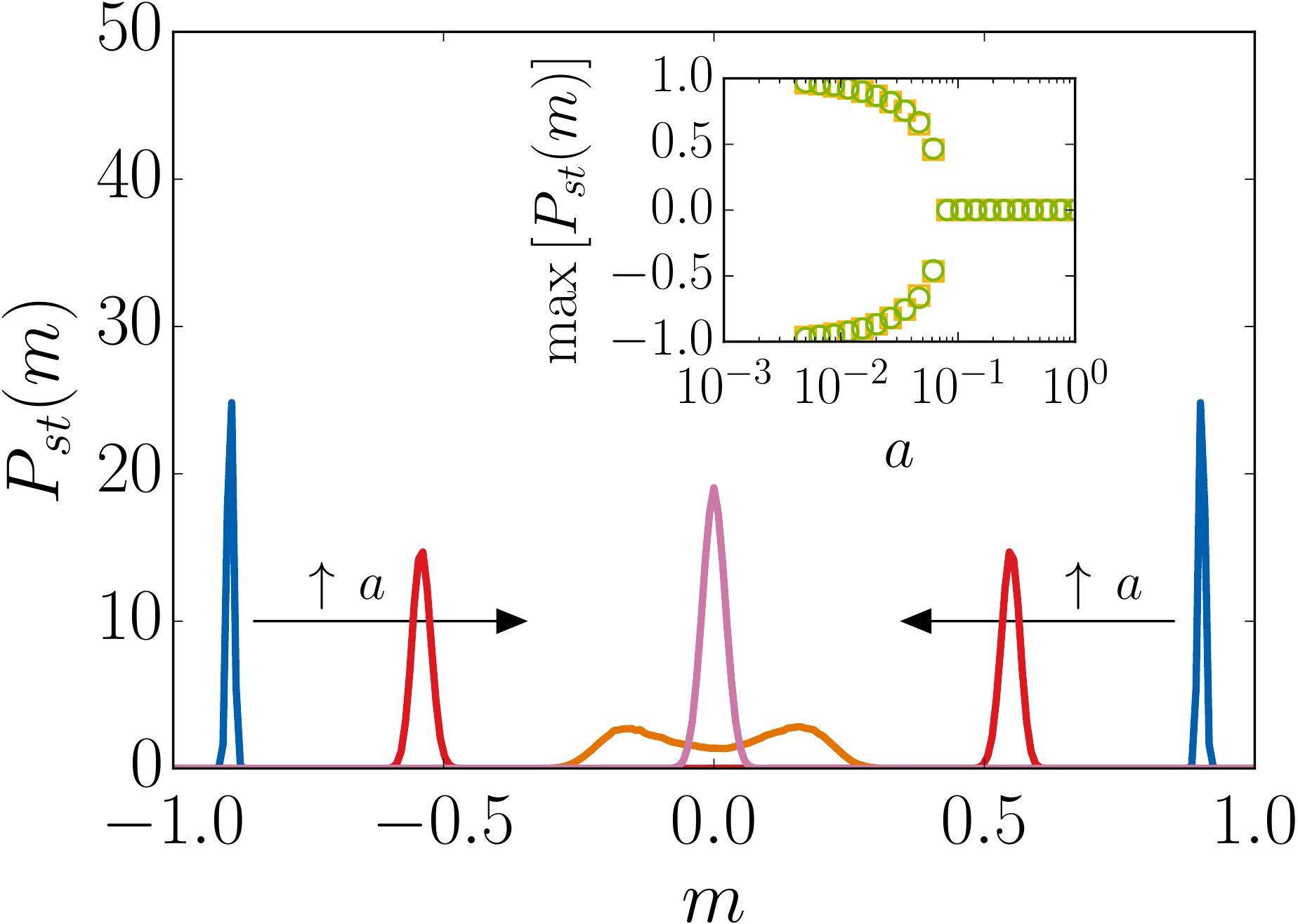}
\caption{Simulations of the noisy voter model with aging. In the main figure, the stationary probability as a function of the magnetization, for different values below (bimodal) and above (unimodal) the critical point $ a_c $. In the inset, positions of the maxima as a function of the noise parameter $ a $ for system sizes $ N = 1000 $ and $ N = 2500 $ (data overlap).}\label{fig:Fig2_Aging}
\end{figure}

By simulating the noisy voter model with aging, we see that depending on the value of the noise the stationary distribution of the magnetization can also be unimodal or bimodal, similar to the behavior observed in the standard noisy voter model (Fig.~\ref{fig:Fig2_Aging}). However, the nature of the unimodal-to-bimodal transition (which can be also seen as an order-disorder transition) is very different to the one of the standard noisy voter model (Fig.~\ref{fig:Fig1}). It is now continuous and stable if we modify the system size (see inset of Fig.~\ref{fig:Fig2_Aging}). The stationary values of the magnetization in the bimodal phase are peaked at $ \pm 1 $ (completely ordered state) only when there is no noise in the system. As $a$ increases, the most probable magnetization values symmetrically move toward the center of the interval. In this regime the system is in a partially ordered state, with predominance of one opinion over the other. At $ a = a_c $, the two maxima continuously merge into a single one located at $ m = 0 $, the most disordered state. For noises $ a \geq a_c $ the system remains in the unimodal regime, the larger the noise, the less spread the stationary distribution.

Despite the inherent non-Markovianity introduced by aging, we can still obtain analytical insights about the transition. We first express the global variables $ n $ and $ N - n $, respectively the number of agents in state $ 1 $ and $ 0 $, as a sum of infinite Markovian variables. Thus, $ n = \sum_{\tau=0}^{\infty} n_{\tau}^+ $ and $ N - n = \sum_{\tau=0}^{\infty} n_{\tau}^- $, where $ n_{\tau}^+ $ and $ n_{\tau}^- $ correspond to the number of nodes in state $ 1 $ and $0$, grouped by their age $ \tau $ \cite{stark2008decelerating}. In terms of the subpopulations we have four possible outcomes after an update event. The first one is $ \{n_{\tau}^{+} \to n_{\tau}^{+} - 1;\, n_0^{-} \to n_0^{-} +1 \}$, and it represents the case in which an agent of age $\tau$ changes from state $1$ to state $0$, thus resetting its internal time. This process occurs at a rate $\Omega_1$. A similar process in which an agent changes from state $1$ to state $0$ occurs at a rate $\Omega_2$. The third process, $ \{n_{\tau}^{+} \to n_{\tau}^{+} - 1;\, n_{\tau+1}^{+} \to n_{\tau+1}^{+} +1 \}$, occurs when the selected agent is in state $1$ and it is not able to change state, thus increasing her internal age by $1$. This process occurs at a rate $\Omega_3$. There are several reasons for this to occur: A noisy update that does not result in a change of state, the agent can not beat the aging when activating, or the agent overcomes the aging but it copies another node with her same state. The final process, with rate $\Omega_4$, is the equivalent one when the selected agent is in state $0$ and it is not able to change state. The rates corresponding to these transitions read, respectively,
\begin{align}
\label{eq:3}
\Omega_1& = n_{\tau}^+ \left( \frac{a}{2} + \frac{1-a}{2+\tau} \frac{N-n}{N} \right), \notag \\
\Omega_2 & = n_{\tau}^- \left( \frac{a}{2} + \frac{1-a}{2+\tau} \frac{n}{N} \right), \\
\Omega_3 & = n_{\tau}^+ \left( \frac{a}{2} + \frac{(1-a)(1+\tau)}{2+\tau} + \frac{1-a}{2+\tau} \frac{n}{N} \right), \notag\\
\Omega_4 & = n_{\tau}^- \left( \frac{a}{2} + \frac{(1-a)(1+\tau)}{2+\tau} + \frac{1-a}{2+\tau} \frac{N-n}{N} \right). \notag
\end{align}

\begin{figure*}
\minipage{0.33\textwidth}
 \includegraphics[width=\linewidth]{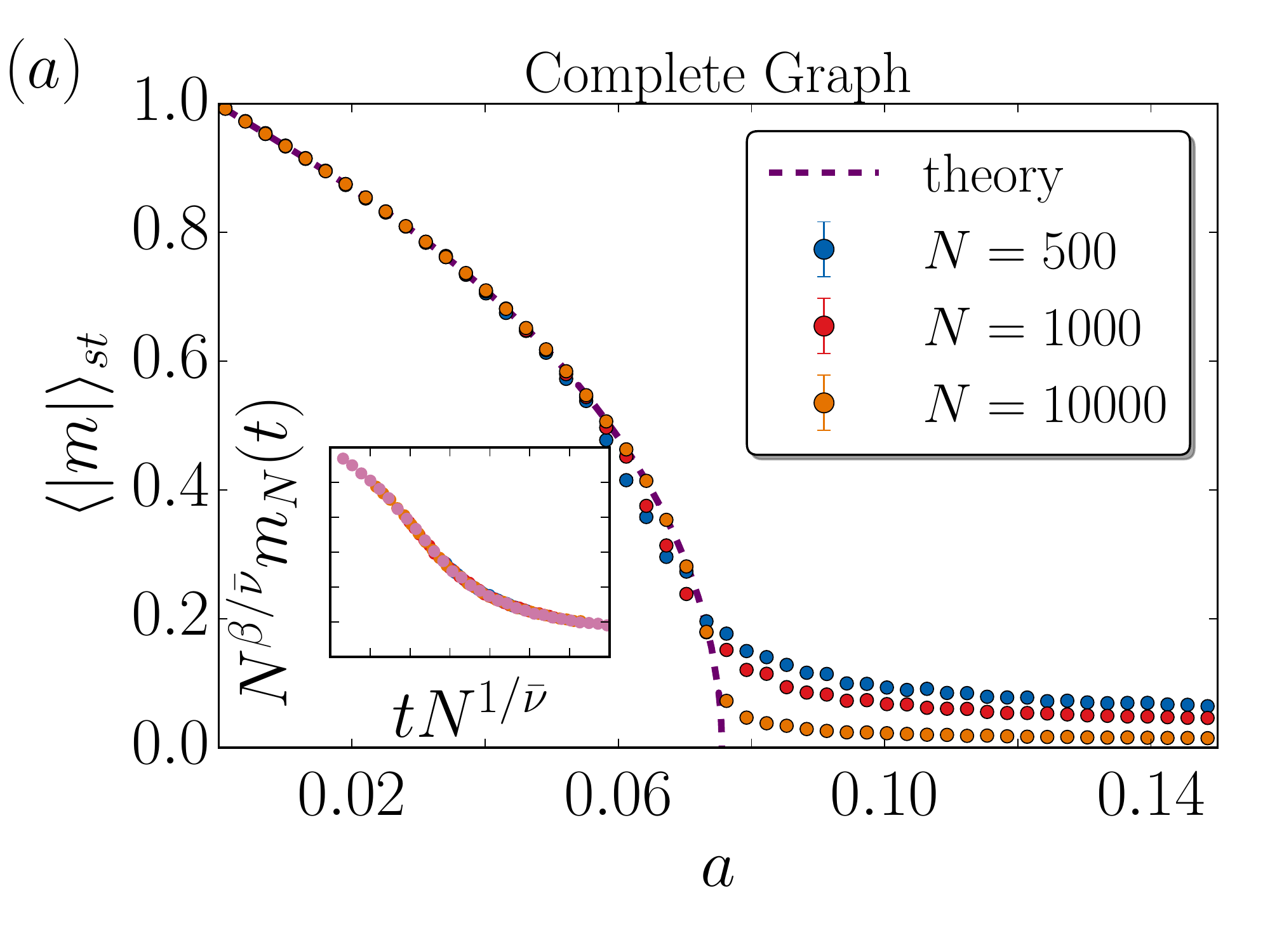}
\endminipage\hfill
\minipage{0.33\textwidth}
 \includegraphics[width=\linewidth]{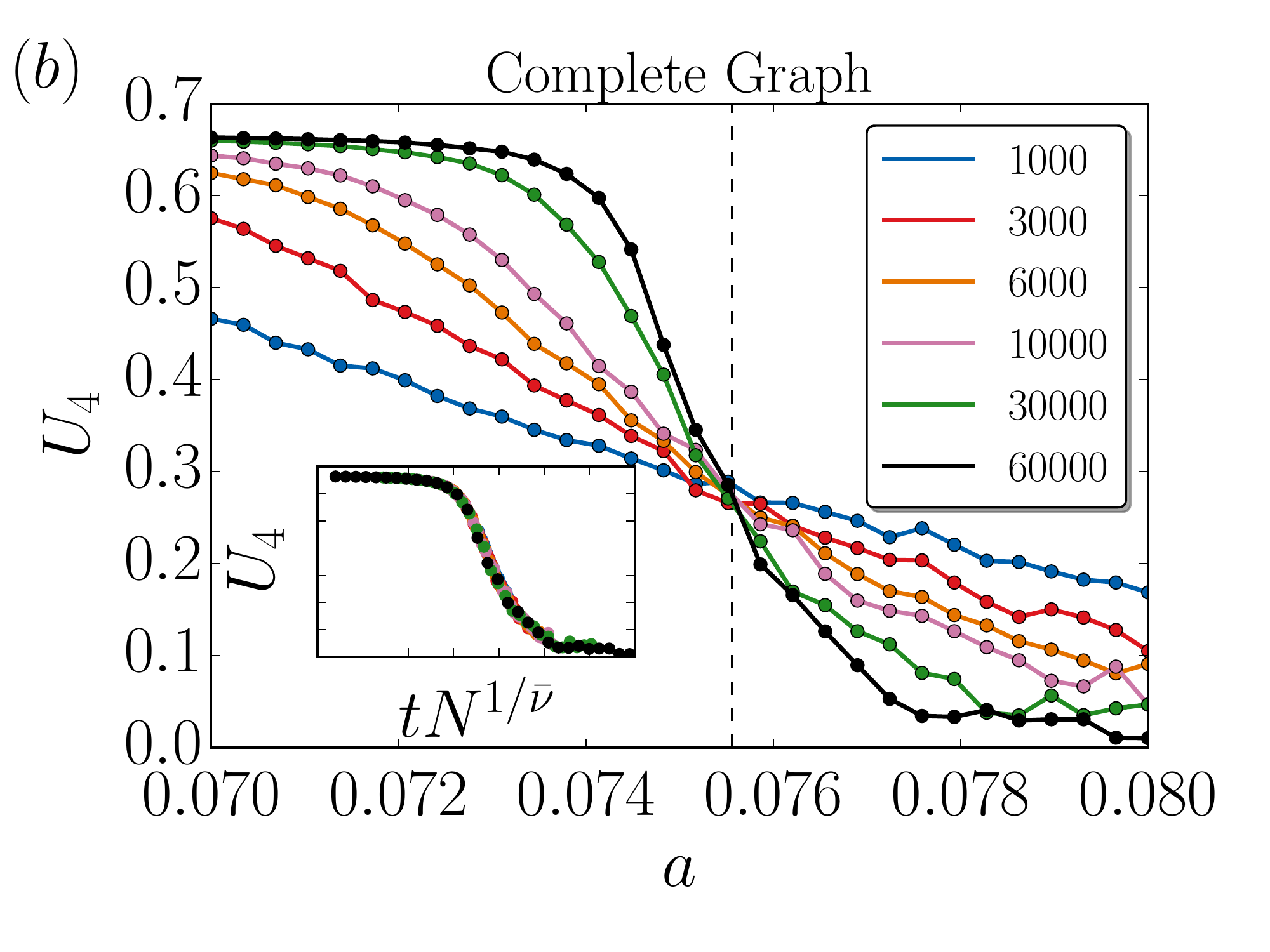}
\endminipage\hfill
\minipage{0.33\textwidth}%
 \includegraphics[width=\linewidth]{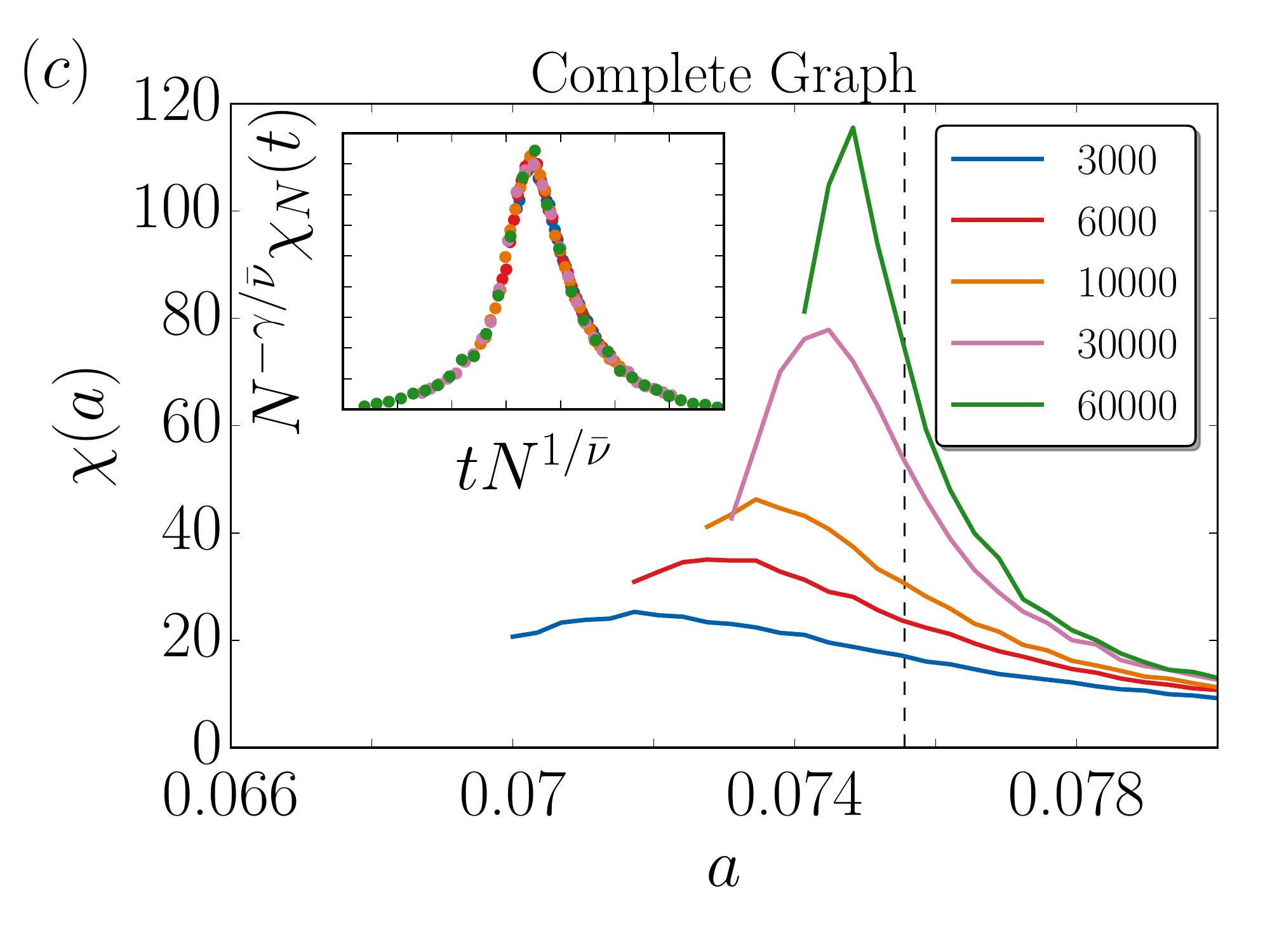}
\endminipage

\minipage{0.33\textwidth}
 \includegraphics[width=\linewidth]{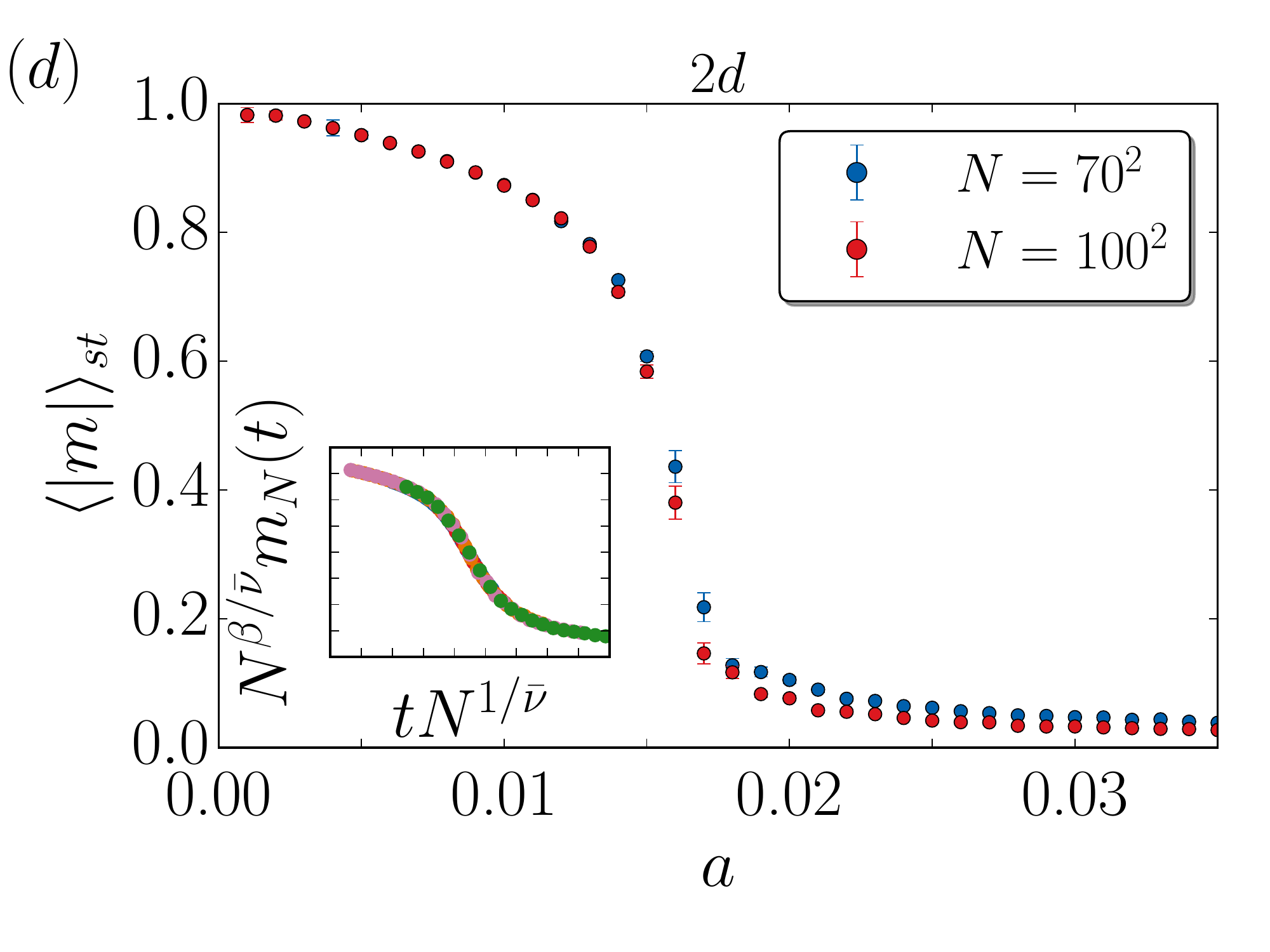}
\endminipage\hfill
\minipage{0.33\textwidth}
 \includegraphics[width=\linewidth]{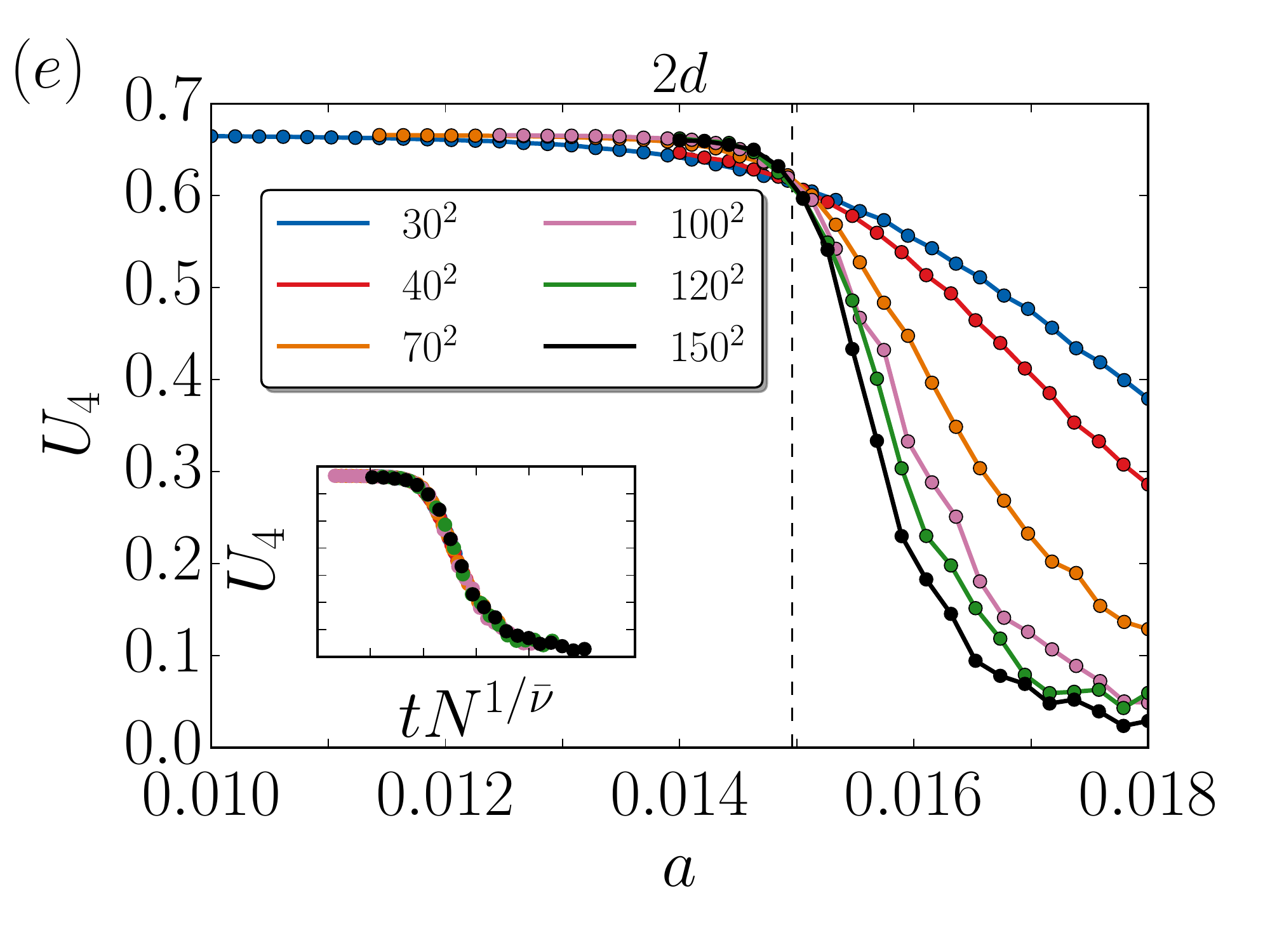}
\endminipage\hfill
\minipage{0.33\textwidth}%
 \includegraphics[width=\linewidth]{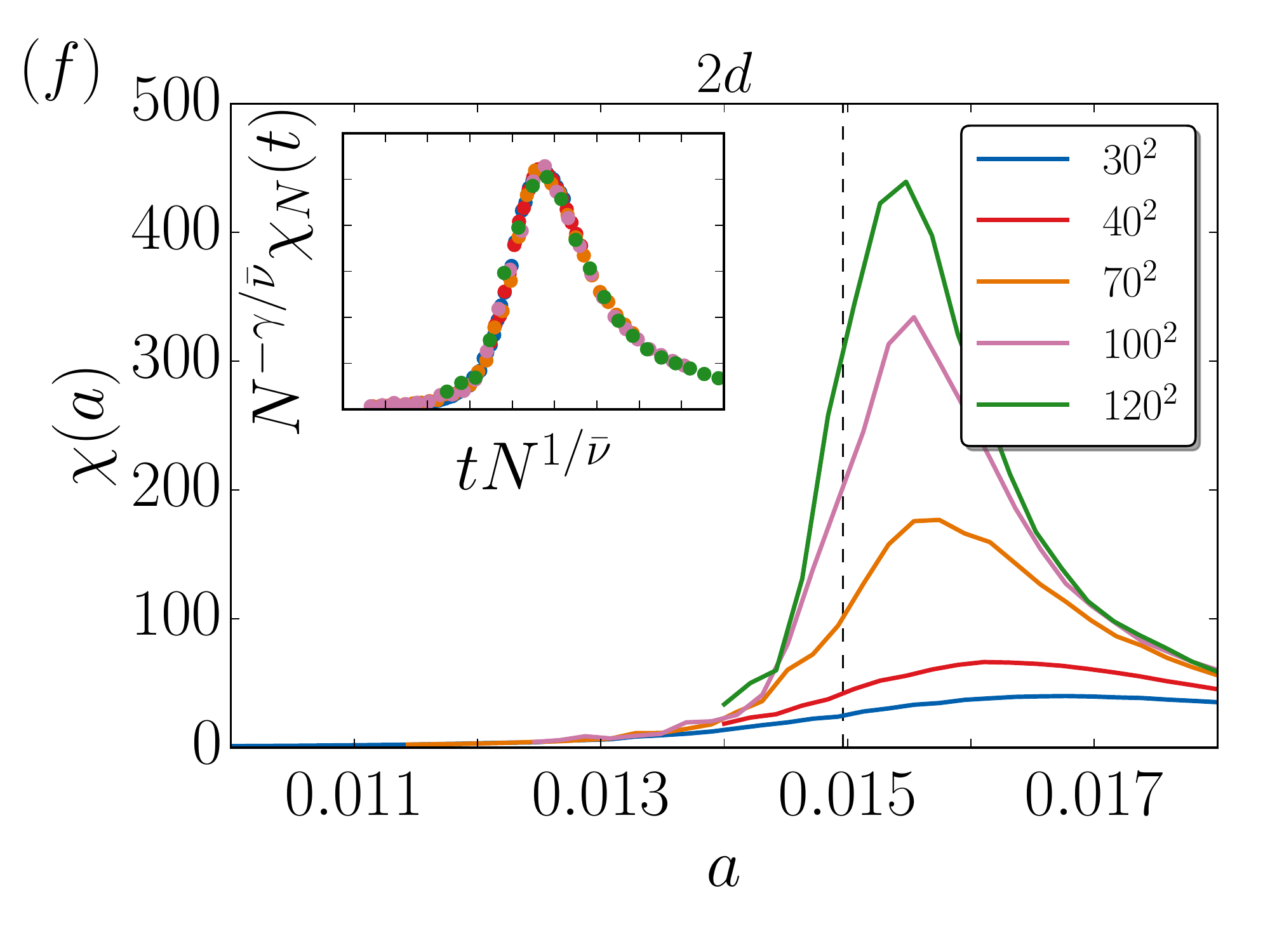}
\endminipage

\caption{($a$, $d$) Stationary magnetization [analytical expression Eq.~(\ref{eq:6}) and simulations], ($b$, $e$) Binder cumulant, and ($c$, $f$) susceptibility for the noisy voter model with aging for different system sizes in a all-to-all interaction [top row] and in a two-dimensional lattice [bottom row]. The insets show the collapses with the corresponding Ising critical exponents \cite{yeomans1992statistical}. The vertical lines are located at the critical point to guide the eye. \emph{Reprinted from Artime et al. Phys. Rev. E 98, 032104 (2018).}
}\label{fig:Fig2}
\end{figure*}

With the transition rates at hand, it is immediate to write an equation for the temporal evolution of the mean value of the different subpopulations $ \langle n_{\tau}^+ \rangle $, $ \langle n_{\tau}^- \rangle $, $ \langle n_0^+ \rangle $ and $ \langle n_0^- \rangle $. These equations form an infinite set of coupled differential equations \cite{artime2018aging}, which is hard to tackle analytically. However, the information given by their stationary solution is enough to characterize the phase transition. We refer the reader to Ref.~\cite{artime2018aging} for further details on how to obtain the conditions of stationarity. Combining these conditions, we are able to obtain the implicit equation for the magnetization $m=2\langle n \rangle_{st} / N-1$:
\begin{equation}
\label{eq:6}
\frac{1+m}{1-m} = \frac{f(a,m)}{f(a,-m)},
\end{equation}
where
\begin{equation}
f(a,m) \equiv \frac{\left(\frac{2}{a}\right)^{\frac{1+(1-a)m}{2-a}}  - 1}{1+(1-a)m}.
\end{equation}
The solution of these equations provide the magnetization as a function of the noise intensity $m(a)$.  Clearly, the solution $m=0$ exists for all values of the noise, but its stability changes: It is a stable solution for $ a > a_c $ and unstable for $a < a_c$. In the regime $ a < a_c $ two new stable and symmetric solutions continuously appear, corresponding to the two ferromagnetic branches. There are no closed analytical expressions but they can be obtained by numerical means. Regarding the critical value $a_c$, it can be obtained by imposing that the derivatives with respect to $ m $ on the two sides of Eq.~\eqref{eq:6} coincide when evaluated at $ m=0 $. After simple but lengthy algebra, one obtains the equation for the critical point,
\begin{equation}
\label{eq:7}
\frac{(2-a_c)^2}{1-a_c} = \log \left(\frac{2}{a_c} \right) \left( 1 - \left( \frac{a_c}{2}\right)^{\frac{1}{2-a_c}} \right)^{-1},
\end{equation}
which gives $ a_c = 0.07556 ... $. We plot in Fig.~\ref{fig:Fig2}$(a)$ the average magnetization $\langle |m| \rangle_{\rm st}$ obtained from numerical simulations in the all-to-all scenario for different system sizes, together with the numerical solution of Eq.~\eqref{eq:6}. We find an excellent agreement between the analytical expression and the simulations.

A Taylor expansion of Eq.~\eqref{eq:6} around $a_c$ yields $ m \sim | a - a_c |^{\beta} $ with $ \beta = 1/2 $ for $a<a_c$ a, which coincides with the $ \beta $-exponent of the Ising universality class in the all-to-all interaction. Hence, this class represents a potential candidate for encompassing also the noisy voter model with aging. To test the accuracy of the prediction for the critical point $ a_c $ we use the Binder cumulant \cite{binder1981finite} $ U_4(a) = 1 - \langle m^4 \rangle_{\rm st} / 3 \langle m^2 \rangle_{\rm st} ^2 $, being $\langle m^{n} \rangle_{st} $ the $ n $-th moment of the magnetization in the stationary state. The critical noise is given by the crossing point of the cumulant curves computed for different system sizes, see Fig.~\ref{fig:Fig2}$(b)$, which turns out to be $ a_c = 0.0753(6)$, compatible, within error bars, with the theoretical prediction. We can also analyze the susceptibility computed as the normalized fluctuations of the magnetization in the stationary state: $\chi(a) = N (\langle m ^2 \rangle_{\rm st} - \langle m \rangle_{\rm st} ^2 )$. It is well-known that at the critical point the susceptibility diverges with the system size. We confirm this tendency in Fig.~\ref{fig:Fig2}$(c)$ by showing the susceptibility for different system sizes.

Aiming at further understating the nature of the phase transition, we can find other critical exponents by employing techniques of finite-size scaling to the curves of the magnetization, the Binder cumulant and the susceptibility. It is important to remark that we cannot define either an internal energy or a specific heat, since we are dealing with an out of equilibrium model that does not possess a Hamiltonian, so not all critical exponents are meaningful for our system. The scaling hypothesis predicts that close to the critical point (i.e., $ t \equiv 1 - a/a_c \to 0 $) the magnetization behaves as $m = N^{-v} f_{1} (t\, N^u)$ and the susceptibility as $\chi = N^w \, f_{2} (t\, N^u)$ with $ v = \beta/(d  \nu) $, $ u = 1/ (d \nu) $ and $ w = \gamma/(d \nu) $, where $ d $ is the dimension in which the system is embedded and $ \gamma $ and $ \nu $ are the critical exponents for the susceptibility and the correlation length, anbd $f_18z)$ ande $f_2(z)$ are scaling functions. Above the critical dimension $d_c$ the exponents become mean field, and they do not depend on the dimensionality anymore. Compactly, we write that $ v = \beta/\bar{\nu} $, $ u = 1/ \bar{\nu} $ and $ w = \gamma/\bar{\nu} $ where $\bar{\nu} = d_c \, \nu $ for $ d \geq d_c $, otherwise $\bar{\nu} = d \, \nu $ \cite{deutsch1992optimized}. Since the topology of the curves that we want to collapse is the all-to-all interaction, we certainly know that we are above the critical dimension. Therefore the finite-size scaling brings information on $ \bar{\nu} $, i.e., the only way to find the actual value $ \nu $ is to know the critical dimension $ d_c $ in advance.

We analytically proved that $ \beta = 1/2 $ for the noisy voter model with aging in all-to-all interactions. This value coincides with the mean-field exponent of the Ising model, so this universality class is a reasonable candidate for our model. In the mean-field regime of this class $ \gamma = 1 $, $ \bar{\nu} = 2 $ and the critical dimension is $ d_c = 4 $. By collapsing the magnetization, the Binder cumulant and the susceptibility (insets of Figs.~\ref{fig:Fig2}$(a)$--\ref{fig:Fig2}$(c)$) using an appropriate rescaling of the axes, we confirm that $\beta$, $ \gamma$, and $ \bar{\nu}$ take these values, although we cannot establish $d_c$ from an all-to-all framework as discussed. To proceed, we can compute the same quantities in lower dimensions and collapse the curves using the corresponding Ising critical exponents assuming $ d_c = 4 $. In case the collapses neatly overlap, we can conclude that the Ising universality class is a solid candidate for the noisy voter model with aging. These curves and their collapses are shown in Figs.~\ref{fig:Fig2}$(d$--$f)$, for a lattice of dimension $ d = 2 $ . The same analyses for lattices of $ d = 3 $ and $ 4, $ and Erd\H{o}s--R\'{e}nyi networks, which have an effective infinite dimensionality, can be found in \cite{artime2018aging}. The overlapping is excellent, ratifying, thus, that our system is compatible with the universality class of the Ising model.

\begin{figure}
 \includegraphics[width=0.8\linewidth]{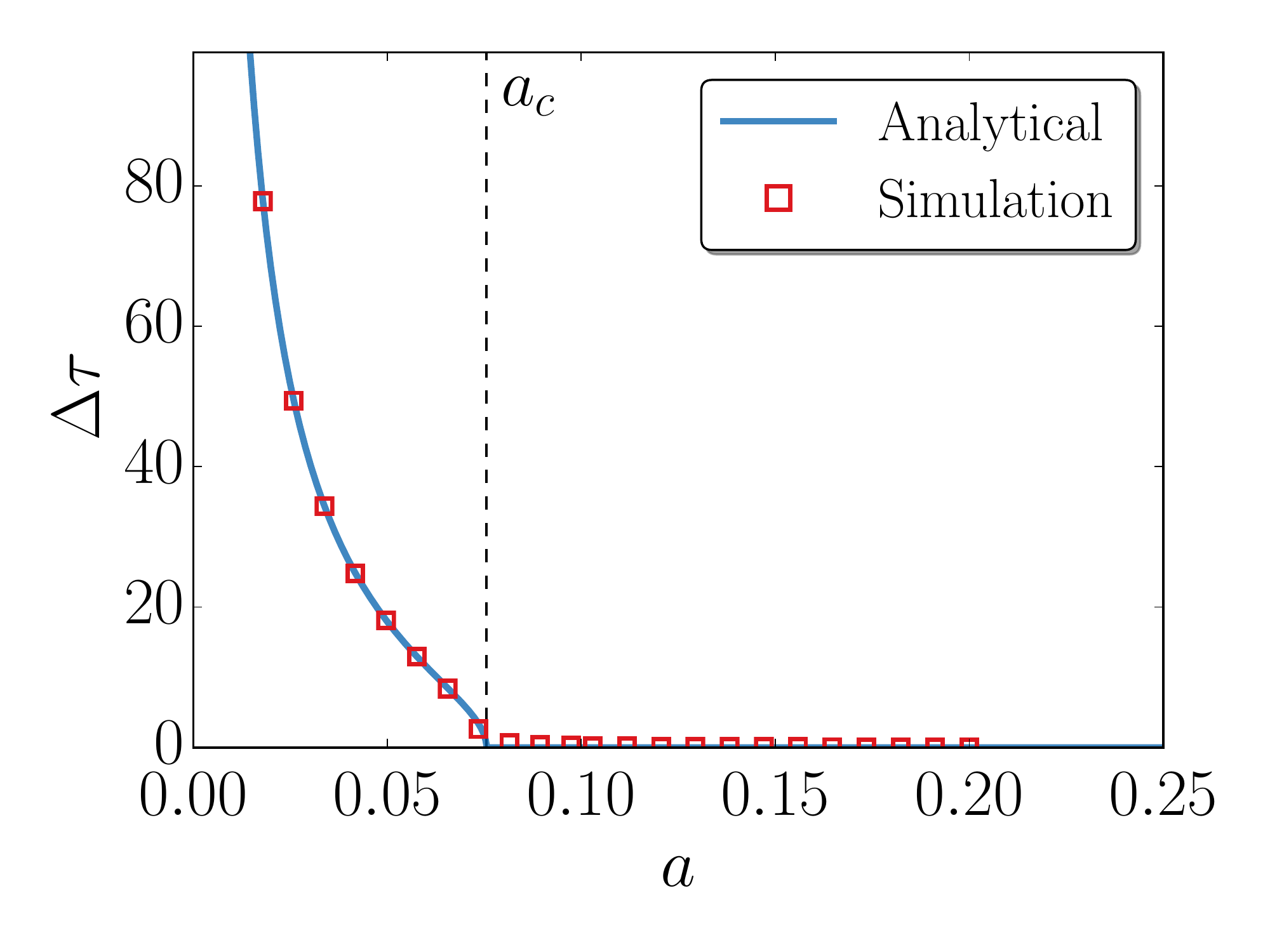}
\caption{Difference of mean internal times $ \Delta \tau = |\overline{\tau}_{\rm M} - \overline{\tau}_{\rm m}| $. The points are obtained from simulations, and the solid lines are obtained from theory, see Eq.(\ref{eq_tau}).}\label{fig:Fig4}
\end{figure}

In the Ising phase transition, a spontaneous symmetry breaking occurs in the order parameter (the magnetization). When aging is at play we can offer an explanation about the symmetry breaking in terms of the internal times too. Initially all nodes have no age, but how are these internal times distributed in the stationary state? Above the critical noise, in the disordered (paramagnetic) phase, the dynamics is driven by noise and no state dominates over the other. Thus, we can expect that all nodes share similar age. Below the critical point, in the ordered (ferromagnetic) phase, there is a non-vanishing value of the magnetization, i.e., a predominance of one state over the other. In this regime, noisy updates are less frequent than the pairwise ones. We then expect an asymmetry in the age distribution of nodes: When there is a global majority opinion, nodes holding this opinion will flip less often than those of the minority, and consequently they will be older, on average. We can quantify this qualitative argument by employing the mean internal time of the agents.  The mean internal time of nodes in state $1$ can be computed as
\begin{equation}
\label{eq_tau}
\overline{\tau}_{+} = \frac{\sum_{\tau} \tau \langle n_{\tau}^{+}\rangle}{\sum_{\tau} \langle n_{\tau}^{+}\rangle} = \frac{1+(1+m) \frac{1-a}{a} \left( \frac{2}{a}\right)^{\frac{1+(1-a)m}{2-a}}}{\left(\frac{2}{a}\right)^{\frac{1+(1-a)m}{2-a}}  - 1}.
\end{equation}
The expression for nodes in state $ 0 $ is obtained substituting $m$ by $-m$. Let $\overline{\tau}_{\rm M} $ be the mean age of the majority population, where the index M is $ + $ or $-$ depending on which opinion dominates, and let $\overline{\tau}_{\rm m}$ be the mean age of the minority population.
We plot in Fig.~\ref{fig:Fig4} their difference $\Delta \tau = |\overline{\tau}_{\rm M} - \overline{\tau}_{\rm m}| $, finding a perfect match between theory and simulations. In the region $ a < a_c $ the asymmetric aging in the populations is evident because of $\Delta \tau \neq 0$. As the system gets closer to the critical noise $\Delta \tau$ vanishes, and it keeps being $0$ in the disordered phase $ a > a_c $. Thus, the difference of the mean internal time of the population gives us a characterization of the phase transition and can be used as an alternative order parameter.

\section{Summary and Conclusions}
\label{sec:summary-and-conclusions}

Simple models of social behavior are useful to isolate basic mechanisms of social interaction and to determine their emergent consequences at a global level and under different conditions. They also provide understanding of different macrolevel behaviors when different mechanisms are introduced in a step by step process. We have discussed in this paper the consequences of a basic mechanism of social learning, namely imitation or herding. When imitation is imperfect, due to individual idiosyncratic choices, there is a transition at a global level from a herding to no a non-herding state. In a first well known description of this phenomenon, this transition disappears for a large number of agents, so that in this limit global herding disappears. We have shown that considering either of two additional mechanisms, nonlinear interactions or aging, this transition becomes robust and well defined for large systems. Nonlinear interactions imply that the probability that an agent changes state is proportional to a power $\alpha$ of the density of her neighbors in the opposite state. Linear interactions correspond to $\alpha=1$. This introduces a second parameter in the model, the first one measuring the proportion of herding to idiosyncratic interactions. In the parameter space $(a,\alpha)$, in an all-to-all interaction as well as in different complex networks, we find three possible phases separated by transition lines and a tricritical point: Unimodal or non-herding phase, bimodal herding phase in each of the possible states or trimodal phase with coexistence of the herding and non-herding phases. On the other hand, aging introduces a non-Markovian dynamics with memory effects in such a way that interactions among agents do not occur at a constant rate, a rather standard implicit hypothesis in most models of interacting agents in Computational Social Sciences. The aging mechanism implies that the longer is the persistence time of an agent in a given state, the less probable is that the agent attempts to update her state by an interaction. We have shown, by analytical calculations in all-to-all interactions and by numerical simulations in lattices of different dimensions and in complex networks, that aging induces a herding to non-herding transition that belongs to the Ising universality class. The transition is associated with a spontaneous symmetry breaking in which the agents in one of the two equivalent states becomes older, in the sense of having a larger mean internal or persistence time.

\section*{Acknowledgements}

Partial financial support has been received from the Agencia Estatal de Investigacion (AEI, Spain) and Fondo Europeo de Desarrollo Regional under Project ESOTECOS Project No. FIS2015-63628-C2-2-R (AEI/FEDER,UE) and the Spanish State Research Agency, through the Maria de Maeztu Program for units of Excellence in R\&D (MDM-2017-0711). A.F.P. acknowledges support by the Formacion de Profesorado Universitario (FPU14/00554) program of Ministerio de Educacion, Cultura y Deportes (MECD) (Spain).

\bibliography{bibliografia.bib}

\end{document}